\documentclass[times, twoside]{zHenriquesLab-StyleBioRxiv}
\usepackage{blindtext}
\usepackage{lipsum}
\usepackage{enumitem}
\usepackage{booktabs}
\makeatletter
\let\cpt@beforestartdoc\relax
\makeatother

\leadauthor{Leire Benito-Del-Valle} 

\begin{document}

\title{Is Synthetic Image Augmentation Useful \\for Imbalanced Classification Problems? \\Case-Study on the MIDOG2025 \\Atypical Cell Detection Competition}
\shorttitle{Synthetic Image Augmentation for MIDOG 2025}

\author[1,2]{Leire Benito-Del-Valle}
\author[1]{Pedro A. Moreno-Sánchez}
\author[1,2] {Itziar Egusquiza}
\author[1]{Itsaso Vitoria}
\author[1,2]{\\Artzai Picón}
\author[1]{Cristina López-Saratxaga}
\author[1]{Adrian Galdran}

\affil[1]{TECNALIA, Basque Research and Technology Alliance (BRTA), Parque Tecnol\'{o}gico de Bizkaia, C/ Geldo. Edificio 700, E-48160 Derio - Bizkaia (Spain)}
\affil[2]{University of the Basque Country, Plaza Torres Quevedo, 48013 Bilbao (Spain) }

\maketitle

\begin{abstract}
The MIDOG~2025 challenge extends prior work on mitotic figure detection by introducing a new Track~2 on \emph{atypical mitosis classification}. This task aims to distinguish normal from atypical mitotic figures in histopathology images, a clinically relevant but highly imbalanced and cross-domain problem. We investigated two complementary backbones: (i) ConvNeXt-Small, pretrained on ImageNet, and (ii) a histopathology-specific ViT from Lunit trained via self-supervision. To address the strong prevalence imbalance (9{,}408 normal vs.\ 1{,}741 atypical), we synthesized additional atypical examples to approximate class balance and compared models trained with real-only vs.\ real+synthetic data. Using five-fold cross-validation, both backbones reached strong performance (mean AUROC $\approx$95\%), with ConvNeXt achieving slightly higher peaks while Lunit exhibited greater fold-to-fold stability. Synthetic balancing, however, did not lead to consistent improvements. On the organizers’ preliminary hidden test set---explicitly designed as an out-of-distribution debug subset---ConvNeXt attained the highest AUROC (95.4\%), whereas Lunit remained competitive on balanced accuracy. These findings suggest that both ImageNet and domain-pretrained backbones are viable for atypical mitosis classification, with domain-pretraining conferring robustness and ImageNet pretraining reaching higher peaks, while naive synthetic balancing has limited benefit. Full hidden test set results will be reported upon challenge completion.
\end{abstract}

\begin{keywords}
Atypical Mitosis classification | Histopathology | Domain shift | Data imbalance | Synthetic augmentation | MIDOG 2025
\end{keywords}

\begin{corrauthor}
adrian.galdran@tecnalia.com
\end{corrauthor}

\section*{Introduction}
Counting mitotic figures is a long-standing prognostic marker in surgical pathology, but automated analysis is challenged by substantial domain shifts across tumor types, scanners, labs, and staining protocols. 
The MIDOG series of challenges was created to stress-test robustness and generalization in this setting; the 2025 edition adds a dedicated track for subtyping mitoses into normal vs atypical, a clinically relevant distinction linked to tumor aggressiveness . [MIDOG 2025 Nature]

In the context of the MIDOG 2025 competition\cite{ammeling_mitosis_2025}, Track 2 (Atypical Classification) formalized the aforementioned task as binary patch-level classification of mitotic figures. 
Training data was released as a set of 128x128 crops derived from the MIDOG++ dataset [X], spanning nine domains (unique combinations of tumor type, species, scanner, and lab), with 10,191 normal and 1,748 atypical annotations. 
Evaluation emphasizes prevalence-robust performance via balanced accuracy (BA), reflecting that atypical mitoses occur at much lower rates ($\approx$20\%) than normal ones. 
The organizers highlighted four core challenges: (1) severe class imbalance, (2) high intra-class variability, (3) subtle inter-class differences, and (4) pronounced cross-domain shifts. 

In order to address the above challenges, in this work, we study whether synthetic image augmentation targeted at the minority category can mitigate class imbalance. 
We pair this with a model zoo spanning both ImageNet-pretrained backbones and domain-pretrained representations (e.g., histopathology-tuned ViT/DiNO variants), and we compare training with real images only versus with synthetic-balanced positives. 
Candidate models are selected by cross-validation and submitted to the challenge platform for blind testing on hidden domains. 
Our aim is to provide a simple, reproducible reference for when and how synthetic positives help (or fail to help) under the distribution shifts emphasized by MIDOG 2025.
Note that previous comprehensive studies have been recently presented benchmarking diverse capabilities of large histopathology image analysis models \cite{marza_thunder_2025,vitoria_benchmark_2025}, like few-shot learning ability, or robust segmentation performance, but here we focus on their use as initialization for common supervised training of models when data of minority categories is scarce.

Our contributions are as follows:
(i) A systematic assessment of minority-class synthesis for atypical mitosis classification under cross-domain shift.
(ii) A comparative study of ImageNet vs. domain-pretrained encoders trained on real-only vs. synthetically balanced data.
(iii) A compact, competition-ready pipeline for selection and submission of robust classifiers on MIDOG 2025 - mitosis classification task.

\section*{Material and Methods}

\begin{figure*}[t]
\centering
\begin{minipage}{\textwidth}
  \centering
  \includegraphics[width=0.18\textwidth]{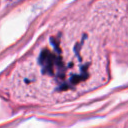}\hfill
  \includegraphics[width=0.18\textwidth]{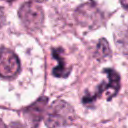}\hfill
  \includegraphics[width=0.18\textwidth]{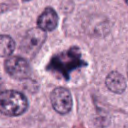}\hfill
  \includegraphics[width=0.18\textwidth]{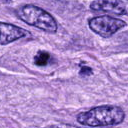}\hfill
  \includegraphics[width=0.18\textwidth]{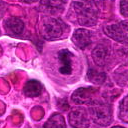}
\end{minipage}

\vspace{1em}

\begin{minipage}{\textwidth}
  \centering
  \includegraphics[width=0.18\textwidth]{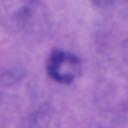}\hfill
  \includegraphics[width=0.18\textwidth]{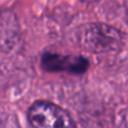}\hfill
  \includegraphics[width=0.18\textwidth]{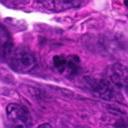}\hfill
  \includegraphics[width=0.18\textwidth]{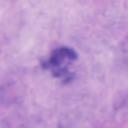}\hfill
  \includegraphics[width=0.18\textwidth]{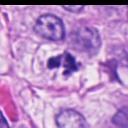}
\end{minipage}
\caption{Top: Real atypical mitotic figure examples. Bottom: Synthetically generated atypical mitotic figures, see Section \ref{syn}.}
\label{fig_real_vs_synth}
\end{figure*}

\subsection{Problem and Data}
The task at hand in MIDOG 2025 is atypical mitotic figure classification, but we first approached the preliminary goal of synthesizing atypical mitotic cells in order to correct the class imbalance of the original dataset. 
The training of the generative model was conducted in two stages. 
To pretrain our generative model for synthetic mitotic figures, we leveraged two large-scale canine histopathology datasets with exhaustive mitotic figure annotations. 
The \textbf{MITOS\_WSI\_CMC} dataset~\cite{aubreville_completely_2020} comprises 32 whole slide images (WSIs) of canine mammary carcinoma, fully annotated for mitotic figures. 
From this resource we extracted approximately 14,000 mitotic figure patches. 
The \textbf{MITOS\_WSI\_CCMCT} dataset~\cite{bertram_large-scale_2019} contains 350 WSIs of canine cutaneous mast cell tumors, again with comprehensive mitotic figure annotations. 
From this dataset we extracted around 42,000 mitotic figure patches. 
In both cases, no distinction is made between typical and atypical mitoses, as such a labelling does not exist for these cohorts. 
These collections nevertheless provide a large, diverse pool of mitotic morphologies that we found useful for pretraining our synthetic generator.

For both the training of the synthetic generator in a \emph{typical vs.\ atypical} setting and for the development of our classification models, we used the dataset released for \textbf{MIDOG~2025, Track~2}~\cite{weiss_2025_15188326}. 
This dataset consists of $128{\times}128$ crops centered on mitotic figures collected from MIDOG++ and partner cohorts. 
Each crop is annotated as either \emph{normal} or \emph{atypical}. 
The dataset covers nine domains, defined by distinct combinations of tumor type, species (human/veterinary), scanner, and laboratory. 
In total, the training set contains 10,191 normal and 1,748 atypical mitoses. 
This dataset thus serves both as the ground truth for classifier training and as the reference distribution for evaluating our synthetic augmentation strategy. 
For visual reference, Fig. \ref{fig_real_vs_synth} shows in the top row five randomly sampled atypical mitotic figures from the MIDOG-2025 dataset, and five synthetic figures below them.
Regarding classifier evaluation data, the official competition score for MIDOG~2025 Track~2 is computed on a hidden test set of 120 cases spanning 12 tumor types (human and veterinary), including two types not present in previous MIDOG editions.

\subsection{Synthetic Image Generation}\label{syn}
There has been recent interest in the generation of synthetic data to boost predictive model performance in histopathology image analysis~\cite{benito-del-valle_unleashing_2025,banerjee_chromosome_2025}.
In this work, we propose using a diffusion model to generate synthetic mitotic figures. These models consist of two main stages: a forward diffusion process that gradually adds noise to the input until it becomes white noise, and a reverse denoising process that reconstructs data from noise. Inspired by \citet{rombach_high-resolution_2022}, we adopt a latent diffusion approach to reduce computational cost by operating in a compressed latent space. The architecture comprises a Variational Autoencoder (VAE) for mapping between pixel and latent spaces, and a Denoising Diffusion Probabilistic Model (DDPM) that performs generation within the latent space. In particular, we opt for the Diffusion Transformer (DiT) \cite{peebles_scalable_2023} architecture, a latent diffusion model which uses a transformer-based model as the backbone for the denoising process. 

While a large dataset of mitotic figure images is available, only a small subset includes class annotations. Therefore, we propose a two-stage training strategy.  In the first stage, we pretrain the VAE and DDPM on the full, unlabelled dataset to capture domain-specific features of mitotic figures. In the second stage, we fine-tune the model using the annotated subset, conditioning the generation process on class labels to enable class-aware synthesis.

In the pretraining stage, both the VAE and DDPM are trained independently on 128×128 resolution images for 1000 epochs with a batch size of 32, using the AdamW optimizer and a learning rate of 0.0001. During fine-tuning, the VAE is trained for 2000 epochs and the DDPM for 5000, each with a batch size of 8, initialized with the weights obtained during pretraining. Fine-tuning is performed using a five-fold cross-validation strategy on the annotated subset, resulting in five independently trained models. 

To augment and balance the real dataset, we generate synthetic images using the fine-tuned models. Specifically, we produce 20,000 atypical and 10,191 normal samples across the five folds. Each fold contributes a distinct subset of synthetic images, generated from models trained on different partitions of the annotated data.\footnote{Later in development, we discarded all synthetic negative samples and retained only synthetic positives due to preliminary classification results.}

All training and generation procedures were implemented using PyTorch (2.4.0), PyTorch Lightning (2.0.9), and torchvision (0.19.0). The experiments were conducted on an Ubuntu 20.04 server equipped with an NVIDIA TITAN X GPU.

\subsection*{Model Training}
In this section we describe the training of our predictive system in detail, following the guidelines proposed in \cite{bertram_histologic_2025}. 

\begin{table*}[t]
\centering
\renewcommand{\arraystretch}{1.4} 
\setlength{\tabcolsep}{8pt}
\caption{Five-fold cross-validation AUROC (\%) for ConvNeXt-Small (ImageNet-pretrained) and Lunit SSL (histopathology-pretrained), trained on real-only vs. real+synthetic data.}
\vspace{1.6em} 
\label{tab_cv_results}
\begin{tabular}{lccccc c}
\toprule
\textbf{Model} & \textbf{F1} & \textbf{F2} & \textbf{F3} & \textbf{F4} & \textbf{F5} & \textbf{Mean} $\pm$ \textbf{SD} \\
\midrule
\textbf{ConvNeXt-Small \cite{liu_convnet_2022} (real)}          & 93.84 & 94.74 & 93.63 & 95.68 & 95.21 & 94.62 $\pm$ 0.78 \\
\textbf{ConvNeXt-Small \cite{liu_convnet_2022} (real+synth)}    & 93.55 & 94.40 & 93.74 & 95.22 & 94.96 & 94.37 $\pm$ 0.65 \\
\textbf{Lunit SSL \cite{kang_benchmarking_2023} (real)}               & 94.22 & 94.97 & 94.13 & 95.25 & 95.31 & 94.78 $\pm$ 0.50 \\
\textbf{Lunit SSL \cite{kang_benchmarking_2023} (real+synth)}         & 93.83 & 94.49 & 93.96 & 95.08 & 94.78 & 94.43 $\pm$ 0.48 \\
\bottomrule
\end{tabular}
\end{table*}

\begin{itemize}[leftmargin=*, itemsep=0.6em, topsep=0.8em]
  \item \textbf{Implementation details:}
    \begin{itemize}[leftmargin=1.2em, topsep=0pt, itemsep=0pt]
      \item \emph{Task:} Binary patch-level classification of mitoses (normal vs atypical).
      \item \emph{Backbones:} We restricted experiments to two architectures, which showed the strongest performance in our preliminary experiments: (i) ConvNeXt-Small \cite{liu_convnet_2022} (ImageNet-pretrained) and (ii) Lunit ViT-based SSL model pretrained on histopathology slides~\cite{kang_benchmarking_2023}. This comparison is meant to understand the impact in performance of domain specialization in model initialization.
      \item \emph{Heads:} Final linear classifier with 1 logit (binary). For the ViT backbone, we wrap the encoder with a specialized classifier that projects the \texttt{[CLS]} token to a single logit.
      \item \emph{Input size \& preprocessing:} Patches are supplied to the ConNext model at their native resoution of $128{\times}128$, and later normalized with ImageNet mean/std. For the ViT model, we are forced to conduct an initial upsampling to $224\times224$ to comply with architecture specification. 
      \item \emph{Augmentations (train):} Random affine (scale with factor s=0.95-1.25, translation, rotation $\leq30^\circ$), color jittering (brightness/contrast/saturation 0.15, hue 0.05), random sharpness adjustment (0.25, $p=0.5$), plus horizontal and vertical flips with probability $p=0.5$. Validation (and test) uses only resize and normalization.
    \end{itemize}
\end{itemize}
\begin{itemize}[leftmargin=*, itemsep=0.6em, topsep=0.8em]
  \item \textbf{Training protocol:}
    \begin{itemize}[leftmargin=1.2em, topsep=0pt, itemsep=0pt]
      \item \emph{Synthetic balancing:} To mitigate class imbalance, we add 7,667 synthetic atypical patches per fold (synthetic negatives = 0), yielding near 50/50 prevalence. We compare \emph{real\_only} vs. \emph{synth\_balanced} regimes.
      \item \emph{Data splits:} 5-fold cross-validation. Training data for synthetic-balanced models includes real and synthetic positive examples, whereas the validation set only contains real data in both scenarios.
      \item \emph{Batch size and Epchs:} Data is grouped in batches of 16 samples, and models are optimized during 25 epochs in all experiments.
      \item \emph{Learning rate and Optmizer:} $1\!\times\!10^{-4}$ for ConvNeXt, $1\!\times\!10^{-5}$ for ViT, with Cosine-Annealing Learning Rate Restarts every 5 epochs. Weighs optimized with the Nesterov-adam algorithm.
      \item \emph{Loss and Model selection:} Binary Cross-Entropy (BCE).The metric for early stopping and model selection is AUROC (though Balanced Accuracy is the primary challenge metric and we also monitor it for sanity-checking purposes).
    \end{itemize}
\end{itemize}
\begin{itemize}[leftmargin=*, itemsep=0.6em, topsep=0.8em]
  \item \textbf{Evaluation protocol:}
    \begin{itemize}[leftmargin=1.2em, topsep=0pt, itemsep=0pt]
      \item \emph{Primary metric:} Balanced Accuracy (BA), as mandated by the challenge. This is used to select our final model for test submission.
      \item \emph{Cross-validation:} Best checkpoint per fold selected by validation AUROC; results averaged across folds and ensembled for submission.
      \item \emph{Hidden test sets:} 
        \begin{itemize}
          \item \emph{Preliminary debug test:} A small set (four tumor types not in the final test) is available for functionality checks. Organizers warn that performance here is not predictive of the final test set.
          \item \emph{Final hidden test:} 120 cases from 12 unseen tumor types (human and veterinary). Only one submission permitted. Results pending.
        \end{itemize}
    \end{itemize}

  \item \textbf{External data usage:}
    \begin{itemize}[leftmargin=1.2em, itemsep=0.6em, topsep=0.8em]
      \item No external datasets used. All training based solely on official MIDOG~2025 Track~2 data. Synthetic positives are generated internally.
    \end{itemize}
\end{itemize}
\begin{itemize}[leftmargin=*, itemsep=0.6em, topsep=0.8em]
  \item \textbf{Reproducibility:}
    \begin{itemize}[leftmargin=1.2em, topsep=0pt, itemsep=0pt]
      \item \emph{Randomness:} The use of fixed random seeds ensure deterministic sampling; cuDNN set to deterministic mode.
      \item \emph{Software:} Python 3.10, PyTorch 2.7.1, Torchvision, HuggingFace Transformers.
      \item \emph{Hardware:} Training on a single NVIDIA GPU (Quadro RTX 6000).
    \end{itemize}
\end{itemize}

\begin{table*}[t]
\centering
\renewcommand{\arraystretch}{1.4}
\setlength{\tabcolsep}{8pt}
\caption{Preliminary hidden test set results (\%) for ConvNeXt-Small (ImageNet-pretrained) and Lunit SSL (histopathology-pretrained), trained on real-only vs. real+synthetic data.}
\vspace{1.6em} 
\label{tab_prelim_results}
\begin{tabular}{lccccc}
\toprule
\textbf{Model} & \textbf{AUROC} & \textbf{Accuracy} & \textbf{Sensitivity} & \textbf{Specificity} & \textbf{Bal. Acc.} \\
\midrule
\textbf{ConvNeXt-Small \cite{liu_convnet_2022} (real)}       & 95.42 & 88.61 & 88.73 & 88.58 & 88.66 \\
\textbf{ConvNeXt-Small \cite{liu_convnet_2022} (real+synth)} & 94.99 & 88.89 & 87.32 & 89.27 & 88.30 \\
\textbf{Lunit SSL \cite{kang_benchmarking_2023} (real)}            & 93.81 & 86.39 & 85.92 & 86.51 & 86.21 \\
\textbf{Lunit SSL \cite{kang_benchmarking_2023} (real+synth)}      & 94.01     & 88.33     & 88.73     & 88.23     & 88.48     \\
\bottomrule
\end{tabular}
\end{table*}

\section*{Results}
Cross-Validation results and then preliminary test set results.

The cross-validation results in Table~\ref{tab_cv_results} show that both ConvNeXt-Small (ImageNet-pretrained) and the Lunit SSL histopathology-pretrained backbone achieve strong performance in atypical mitosis classification, with mean AUROC values in the 94–95\% range across five folds. ConvNeXt-Small reaches a slightly higher mean AUROC (94.62\%) than Lunit (94.78\%), but at the cost of noticeably higher variability across folds (standard deviation $\approx$0.78 vs. 0.50). 
This suggests that the histopathology-specialized features from Lunit yield more stable performance for in-distribution data. 

Synthetic balancing, which added 7,667 atypical examples to approximate class balance, did not consistently improve results. 
In both backbones, the real-only models performed marginally better than their synth-balanced counterparts (ConvNeXt: 94.62 vs. 94.37; Lunit: 94.78 vs. 94.43), although the differences are within one standard deviation and likely not statistically significant. 
This indicates that while synthetic positives may help counter severe imbalance in principle, their benefit is limited in the presence of sufficiently large real training data, and care is needed to avoid introducing domain artifacts.

Overall, these results highlight that (i) both ImageNet and domain-pretrained models are competitive for atypical mitosis classification, (ii) histopathology-pretrained backbones offer more stable generalization across folds, and (iii) synthetic balancing does not translate into clear cross-validation gains.

Submission to the hidden test set enables stress-testing our models’ robustness against domain shifts.
Unfortunately, at the time of writing, only a small portion of the full test set is made available for performance analysis purposes\footnote{This preprint will be updated with full results once the competition finishes.}.
Results on this preliminary test set are shown in Table~\ref{tab_prelim_results}.
Please note that this dataset is not aligned with the full test data distribution, and thus not predictive of final test performance.
Under this disclaimer, we can see that ConvNeXt-Small achieves the strongest results overall, with AUROC values around 95\% and balanced accuracy close to 89\%.
The Lunit SSL backbone also performs competitively, but trails ConvNeXt by 1–2 percentage points across most metrics.
Interestingly, adding synthetic atypical examples does not provide a consistent improvement: the real-only ConvNeXt model slightly outperforms its synth-balanced counterpart in AUROC, while the synth-trained model compensates with marginal gains in specificity.
For Lunit SSL, preliminary trends suggest that domain-pretrained features are stable though not necessarily superior on this out-of-distribution subset.
Overall, these observations reinforce the cross-validation findings: both ImageNet and domain-pretrained backbones are competitive, ConvNeXt reaches higher peak performance, and straightforward synthetic balancing has limited effect on generalization in this particular problem.
Our final submission was selected to be as the one achieving highest AUROC values in the preliminary test set, i.e. an ensemble of ConvNeXt-Small networks trained on real images only.
\section*{Discussion}
\paragraph{Summary of findings.}
Across five folds, both ConvNeXt-Small~\cite{liu_convnet_2022} and the Lunit SSL ViT~\cite{kang_benchmarking_2023} achieve strong AUROC in the mid–high 90s, with the domain-pretrained backbone showing slightly \emph{more stable} performance (lower fold-to-fold variance), while ConvNeXt tends to reach slightly \emph{higher peaks} on some folds. On the organizers’ preliminary hidden test set—explicitly stated to be domain-mismatched—ConvNeXt narrowly leads on AUROC and Accuracy, whereas Lunit remains competitive on Balanced Accuracy. These trends broadly mirror our cross-validation behavior.

\paragraph{On synthetic balancing.}
Injecting $7{,}667$ synthetic atypical crops to approximate a 50/50 prevalence did \emph{not} produce consistent gains in our setting. Real-only models were marginally stronger or on par in mean AUROC/BA. Two factors may explain this: (i) residual domain gap between synthetic atypicals and real data (style, texture, or morphological priors), and (ii) sufficient effective sample size of real positives once augmentation and five-fold reuse are accounted for. 
More targeted synthesis, e.g. class-conditional style transfer, histopathology-aware diffusion priors, or curriculum mixing by confidence, may unlock larger benefits.

\paragraph{Limitations.}
First, the preliminary hidden test set is small and intentionally out-of-distribution relative to the final evaluation; conclusions from it should be viewed as sanity checks, not benchmarks. 
Second, our synthetic pipeline focused only on minority-class did not explore stain/style harmonization, domain adversarial learning, or explicit domain generalization, all of which may improve atypical subtype robustness. 
Third, we did not quantify the effect of thresholding and calibration on BA in depth (e.g., Platt/temperature scaling or isotonic), which could further stabilize specificity–sensitivity trade-offs.

\paragraph{Future work.}
(i) Explore histology-aware generative augmentation (e.g., diffusion with morphology constraints) that preserves nuclear architecture while varying stain and context; (ii) couple synthesis with style transfer and color augmentation to better match unseen labs; (iii) investigate DG/DA methods (IRM, SWAD, feature whitening) and stain normalization; (iv) incorporate confidence calibration and conformal risk control for threshold selection under BA; and (v) perform ablations on early-stopping criteria (AUROC vs.\ BA) and ensembling strategies across folds/backbones.

\begin{acknowledgements}
A. Galdran was supported by grant RYC2022-037144-I, funded by MCIN/AEI/10.13039/501100011033 and by FSE+. 
In addition, this work was partially funded by the European project SEARCH (\url{https://ihi-search.eu/}), which is supported by the Innovative Health Initiative Joint Undertaking (IHI JU) under grant agreement No. 101172997. The JU receives support from the European Union’s Horizon Europe research and innovation programme and COCIR, EFPIA, Europa Bio, MedTech Europe, Vaccines Europe, Medical Values GmbH, Corsano Health BV, Syntheticus AG, Maggioli SpA, Motilent Ltd, Ubitech Ltd, Hemex Benelux, Hellenic Healthcare Group, German Oncology Center, Byte Solutions Unlimited, AdaptIT GmbH. 
Views and opinions expressed are however those of the author(s) only and do not necessarily reflect those of the aforementioned parties. Neither of the aforementioned parties can be held responsible for them. 
\end{acknowledgements}

\end{document}